\documentclass[preprint2]{emulateapj}
\usepackage{amsmath}
\usepackage{graphicx}
\usepackage{natbib}

\newcommand{\D}{{\cal D}}

\def\gsim{ \lower .75ex \hbox{$\sim$} \llap{\raise .27ex \hbox{$>$}} }
\def\lsim{ \lower .75ex\hbox{$\sim$} \llap{\raise .27ex \hbox{$<$}} }

\shorttitle{GRBs' GeV Emission} \shortauthors{Piran and Nakar}

\begin{document}

\title{On The External Shock Synchrotron Model for GRBs' GeV Emission. }

\author{Tsvi Piran\altaffilmark{1}  }
\affil{Racah Institute of Physics, Edmund J. Safra Campus
Hebrew University of Jerusalem, Jerusalem 91904, Israel}
\and
\author{ Ehud Nakar\altaffilmark{2}}
\affil{Raymond and Beverly Sackler School of Physics and
Astronomy, Tel Aviv University, Tel Aviv 69978, Israel}
\altaffiltext{1}{tsvi@phys.huji.ac.ill}
\altaffiltext{2}{Udini@wise.tau.ac.il}

\begin{abstract}
The dominant component of the  (100 MeV - 50 GeV) GRB emission
detected by LAT starts with a delay relative to the prompt soft
(sub-MeV) gamma-rays and lasts long after the soft component fades.
This has lead to the intriguing suggestion that this high energy
emission is generated via synchrotron emission of relativistic
electrons accelerated by the external shock. Moreover, the limits on
the MeV afterglow emission lead to the suggestion that, at least in
bright GeV bursts the field is not amplified beyond compression in
the shock. We show here that considerations of confinement (within
the decelerating shock), efficiency and cooling of the emitting
electrons constrain, within this model, the magnetic fields that
arise in both the upstream (circum burst) and downstream (ejecta)
regions, allowing us to obtain a direct handle on their values. The
well known limit on the maximal synchrotron emission, when combined
with the blast wave evolution, implies that late photons (arriving
more than $\sim$ 100 s after the burst) with energies higher than
$\sim $10GeV do not arise naturally from external shock synchrotron
and  almost certainly have a different origin. Finally,  even a
modest seed flux (a few mJy) at IR-optical would quench, via
Inverse Compton cooling, the GeV emission unless the magnetic
field is significantly amplified behind the shock. An observation
of a burst with simultaneous IR-optical and GeV emission will
rule out this model.
\end{abstract}

\section{Introduction}
The recent observations of the Large Area Telescope (LAT) on board
of Fermi of   GeV emission (100MeV - 50GeV) from GRBs revealed
an interesting pattern. The GeV emission is delayed relative to the
onset of the prompt MeV emission \citep{Abdo080916C}. It shows a
constant power-law decay long after the prompt emission dies out
\citep{Abdo090902B,GhiselliniEtal10}. While surprising at first, one
may recall that a ''precursor" of these observations was made
already by EGRET that detected an $18$ GeV photon 90 minutes after
the burst in  GRB 940217 \cite{Hurley94} and a rising late GeV
spectral component in GRB 941017 \citep{GonzalezEtal03}. This
pattern suggests that the bulk of the GeV emission arises from an
external shock afterglow \citep{KB09a,KB09b,GhiselliniEtal10}. While
a detectable high energy external shock emission was expected for a
long time \citep{MR94} and it was noted that external shock
synchrotron emission may be the strongest afterglow GeV component
(see. e.g. \citealt{Fanetal08,FanPiran08,ZouEtal09}), the observation
that this may be the dominant GeV emission over the whole
burst, including the prompt phase, were surprising.

Following these observations \cite{KB09a,KB09b} proposed a
revolutionary model  in which they revise a critical component of
the standard external shock scenario. They suggest that there is no
magnetic field amplification beyond the usual shock compression.
Namely,  the downstream (shocked) magnetic field is just $4 \Gamma$
(where $\Gamma$ is bulk Lorentz factor behind the shock) times
the upstream circum burst field.   In doing so they are able to fit
the overall afterglow spectrum (ranging from optical to GeV), as the
low magnetic field in the emitting region quenches the lower energy
emission. Additionally they  get rid of a nagging theoretical
problem - how are the fields amplified \citep{Gruzinov01}?

The magnetic field plays a triple role in the synchrotron-shock
acceleration mechanism. It accelerates the electrons and confines
them to the shock, while they are accelerated and it also controls
the synchrotron emission. A weaker magnetic fields poses two
challenges to the model: cooling and confinement. First, a
comparison of the acceleration and the cooling times sets an
absolute limit on the energy of synchrotron photon in the radiating
fluid frame. Together with the hydrodynamics of the decelerating
blast wave this puts a time dependent limit on the maximal energy of
observed synchrotron photons. Photons above this limit are (almost
certainly) not emitted by external shock synchrotron. Efficient
cooling poses another limit on the model. A significant (though not
dominant) amount of energy is emitted in the GeV emission. This
implies that the emitting electrons must be fast cooling. Otherwise
the system would be inefficient and  the energy requirement
unreasonable. As the cooling takes place mostly in the downstream
region this last  condition  constrains the magnetic field there.
While our original motivation was to examine the ''unamplified"
magnetic field scenario our analysis is more general and we allow
for an amplification factor. We show that the observations of a
significant GeV flux poses strong limits on the downstream magnetic
field. These limits can be translated to limits on the upstream
circum burst field (in the case of no amplification) or on the
amplification factor.

Confinement is most important in the upstream region, where the
magnetic field is weakest. Thus, observations of GeV photons limit
the upstream magnetic field with a weak dependence on field
amplification in the shock. Finally we turn to the influence of
Inverse Compton (IC) cooling on the observed GeV emission.  Given
the strong low energy (IR-optical) radiation fields (from the prompt,
reverse shock and the forwards shock itself), the magnetic field
density should be strong enough in order that IC cooling won't
quench the GeV emission. This sets yet another, independent, limits
on the downstream magnetic field. These considerations shed a direct
light on the magnetic fields which are among the most elusive
parameters of the external shock model. Note that here we derive
constraints assuming that the external shock is adiabatic. If it is
radiative (as suggested e.g., by \citealt{GhiselliniEtal10}) then
the constraints will be more stringent.

We examine in this letter these limits that arise from the GeV
emission. We don't attempt to provide a complete solution to the
whole multiwavelength afterglow. As such our analysis is very
general and it depends only on the assumptions of synchrotron
process and the blast wave hydrodynamics.

\section{The Maximal Synchrotron Energy}

It is well known \citep[e.g.,][]{deJager92,Lyutikov09,Kirk10} that
by equating the synchrotron cooling  to the acceleration rate, one
can obtain an upper bound on the maximal energy of a  synchrotron
photon:
\begin{equation}
h \nu_{Max}'=  \frac{m_e c^2}{\alpha}.  \label{Emax}
\end{equation}
This limit is in the    fluid's  rest frame, denoted by ``$~'~$". $h$ is Planck constant, $m_e$ is the electron rest mass, $c$
is the speed of light  and $\alpha$ is the fine structure constant. The maximal
observed energy is larger by the Lorentz boost factor from the fluid
frame to the observer frame, $\D =[\Gamma(1-\beta\mu)(1+z)]^{-1}$,  where
$\Gamma$ is the Lorentz factor of the fluid and $\mu$ is cosine the
angle between the fluid velocity and the line-of-sight.

If the emission originates from a decelerating external shock then
at any given time the observer receives simultaneously photons
emitted from a range of radii and therefore a range of Lorentz
factors. The photons observed at a time $t$ all satisfy
$t=T-R\mu/c$, where $T$ is the time since the explosion as measured
in the source frame and $R$ is the shock radius at time $T$. In a
circum burst medium with mass density $\rho \propto r^{-k}$ this
condition implies that along the line-of-sight (i.e., $\mu=0$) the
Lorentz boost is $\D_{los}=2\Gamma_{los}/(1+z)$, where
$\Gamma_{los}^2={R_{los}}/{4(4-k)ct}$.  The Lorentz factor from any
other direction,  $\mu$, (keeping $t$ constant)  is larger,  but not
necessarily the boost, which satisfies:
\begin{equation}\label{EQ Boost}
  \frac{\D}{\D_{los}}=\frac{\Gamma}{\Gamma_{los}}
  \left[\frac{2(4-k)}{7-2k+\left({\Gamma}/{\Gamma_{los}}\right)^{\frac{2(4-k)}{3-k}}}\right].
\end{equation}
Maximizing $\D$ for $\Gamma \geq \Gamma_{los}$ shows that in a
constant density medium($k=0$), e.g., interstellar medium (ISM),
$\D_{max} = 1.2 ~\D_{los}$ while in a circumburst wind ($k=2$)
$\D_{max} = \D_{los}$. For a self-similar adiabatic shock that
propagates into ISM with a constant particle density, $n$,
$\Gamma_{los,ISM} = [17 E t^3/16^4 \pi  m_p c^5 n(1+z)^3]^{1/8}$
while in a wind with a mass density $\rho = A r^{-2}$,
$\Gamma_{los,wind} = [9 E t/128 \pi c^3 A(1+z)]^{1/4}$ \citep{BM76}.
Here $E$ is the blast wave kinetic energy, $m_p$ is the proton mass
and $z$ is the burst's redshift.

Using these expressions we find that the maximal observed energy of
a synchrotron photon emitted by an adiabatic external shock during
its decelerating phase is:
\begin{equation}\label{eq hnumax}
h \nu_{max}=
\begin{cases}
9{\rm~GeV~} \left(\frac{E_{54}}{n}\right)^{1/8}
\left(\frac{1+z}{2}\right)^{-5/8} t_2^{-3/8} & {\rm  ISM, }  \\
5{\rm~GeV~} \left(\frac{E_{54}}{A_*}\right)^{1/4}
\left(\frac{1+z}{2}\right)^{-3/4} t_2^{-1/4} & { \rm wind, }
\end{cases}
\end{equation}
where we use the common notations of $t_2 = t/100$,
$E_{54}=E/10^{54}$ in c.g.s. units, etc. $n$ is the ISM density and
$A_*$ is the wind parameter in units of $5 \cdot 10^{-11} {~\rm gr
/cm^{-1}}$. Note that the dependence on the burst's parameters of
this rather general condition is very weak. Moreover, even when the
burst redshift is unknown this limit peaks at $z \sim 1$. 
In principle this limit can be violated by synchrotron emission in a
special magnetic field configuration, for example, if the electrons
are accelerated by  a weak magnetic field but radiate where the
field is strong \citep{Lyutikov09, Kumar10}.  Still, it is unlikely that afterglow
photons $\gg 10 $ GeV are generated by synchrotron radiation.

Such photons were already observed in two bursts. Most notable is
GRB 940217, where an $18$ GeV photon was observed $90$ min after the
burst by EGRET \citep{Hurley94}. An order of magnitude larger than
the limit of Eq. \ref{eq hnumax} at this time, $\approx 2$ GeV.
Recently, LAT detected a $33$ GeV photon $82$ s after the burst from
GRB 090902B \citep{Abdo090902B}. The observation of one or two
photons cannot rule out the external shock synchrotron model for GeV
emission. Nevertheless, these observations set a major difficulty to
this model, especially if observations of late afterglow photons
much above this limit will continue.

\section{Confinement and Cooling}
If the observed GeV emission arises from synchrotron in an
external shock it constrains both the upstream and the downstream
magnetic fields. It constrains the first by the requirement that the
radiating electrons are confined to the system. It also constrains
the latter by the requirement that the radiating electrons  radiate
efficiently and therefore are cooling fast. The limits are
particularly interesting in the case that the downstream field is
not amplified beyond the usual shock compression and $B_{d} = 4
\Gamma B_u$. Such a case is interesting for this scenario because a
rather weak magnetic field is needed in order to suppress the
synchrotron MeV emission of the forward shock \citep{KB09a}.
However, in order to consider the most general case, and since IC
emission may suppress the MeV emission without affecting the GeV
luminosity, we introduce an amplification factor, $f_B \ge 1 $, so
that $B_{d} = 4f_B \Gamma B_u$.  In the following we presents these
constraints using $f_B$ and upstream field, which for abbreviation
we denote simply as $B$.

\subsection{Confinement}

The accelerated electrons have to be confined to the shock region,
otherwise they escape. This sets a second limit on the maximal
synchrotron energy.
\begin{figure}[ht!]
 \centering
 \includegraphics[width=3.8 in]{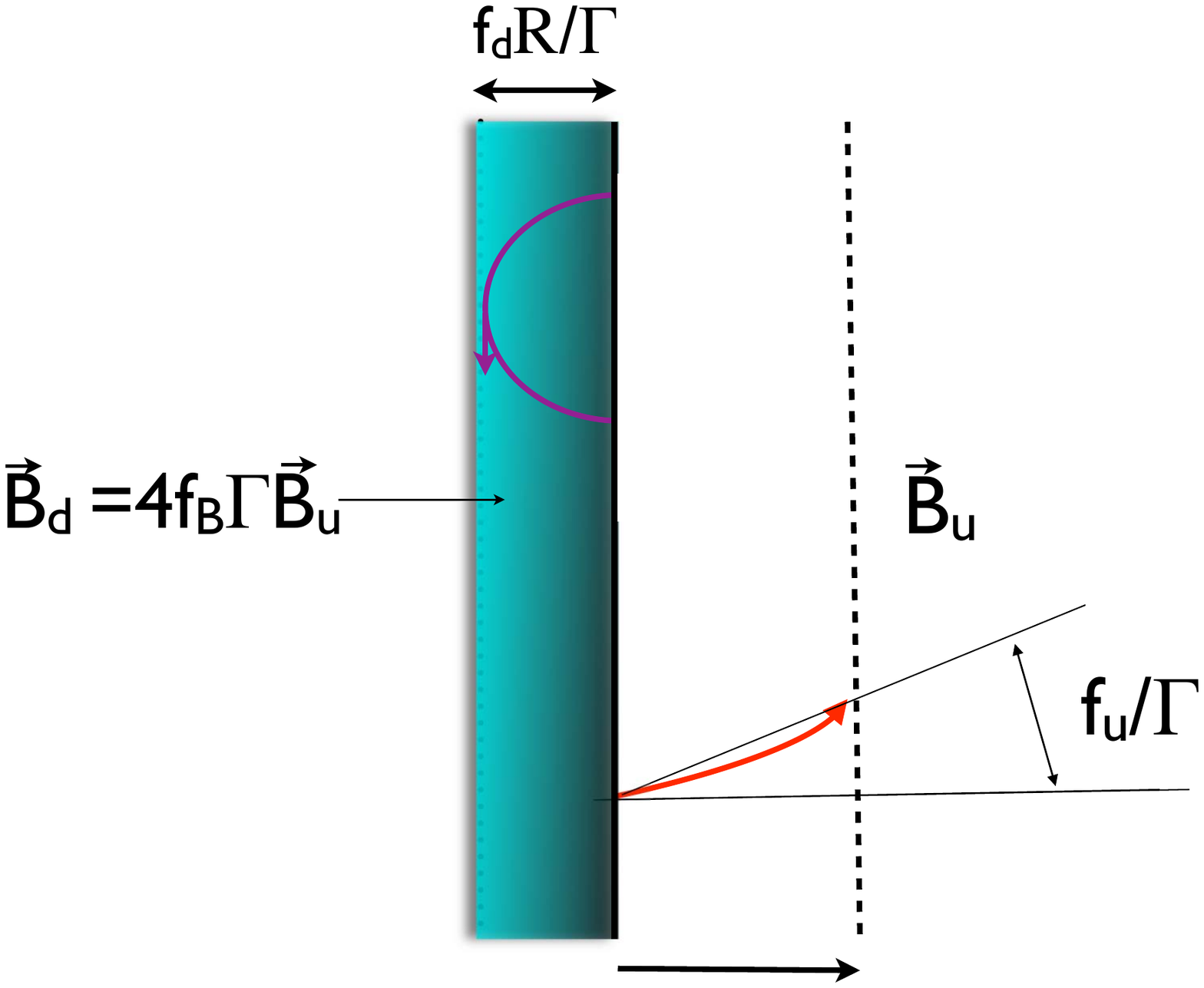}
 \caption{The motion of electrons in the upstream (right - red) and downstream (left - purple). Note that the shock
moves forwards and slows down while the electron gyrates in the upstream magnetic field.  }
 \label{Fig1}
\end{figure}
The confinement criterion on the  Gyro radius in the upstream of
an electron with Lorentz factor $\gamma_e'$  is:
\begin{equation}
R_L =  \frac{\Gamma \gamma_e'  m_e c^2}{e B}  \le f_u \Gamma R ~.
\end{equation}
The external shock slows down while it propagates and
it won't catch the rapid electron by the time it turned $R_L/\Gamma$
but somewhat later (see Fig. \ref{Fig1}). This leads to the confinement factor,  $f_u \le 1 $.
Detailed calculations yield $f_u \approx 1/3$ for an ISM and 1/2 for a wind, we denote
$f_{u,x}=f_u/x$. This last inequality leads to a maximal confined
electron's Lorentz factor (in the shock's frame):
\begin{equation}
\gamma_{conf}' = \frac{ e B} {m c^2 } f_u R ~. \label{conf}
\end{equation}
For confinement in  the downstream the Larmour radius should be
compared with the shock's thickness $f_d R /\Gamma$ where $f_d
\approx 0.1$. The resulting maximal electron Lorentz factor is $4
f_d f_B/f_u$ times larger than the one given by Eq. \ref{conf} for
the upstream. Since the electron has to be confined in both upstream
and downstream  the comparable or smaller  former limit is the
critical one.

The synchrotron frequency of electrons in the downstream is larger
by a factor of $4 f_B$ than in the upstream. Therefore the
confinement upper limit for synchrotron external shock photons is:
\begin{equation}
h \nu_{conf} =  4 f_B \hbar \left( \frac{e B}{m c} \right)^3
\frac{f_u^2 R^2 \Gamma^2 }{(1+z) c^2} ~. \label{syn_conf}
\end{equation}

The average Lorentz factor and the blast wave radius in a constant
density environment that corresponds to an observer time $t$ are
\citep{SPN98}:
\begin{equation}\label{eqgs}
\Gamma = 180 ~ \left(\frac{E_{54}}{n}\right)^{1/8}
\left(\frac{1+z}{2}\right)^{3/8} t_2^{-3/8} ~,
\end{equation}
and
\begin{equation}\label{eqrs}
R = 2 \cdot 10^{17} ~{\rm cm}~ \left(\frac{E_{54}}{n}\right)^{1/4}
\left(\frac{1+z}{2}\right)^{-1/4} t_2^{1/4} ~.
\end{equation}
Substitution of  these expressions into Eqs. \ref{conf} and
\ref{syn_conf} yields:
\begin{equation}
\gamma_{conf}' = 3 \cdot 10^8~ f_{u,\frac{1}{3}} B_{-5}
\left(\frac{E_{54}}{n}\right)^{1/4}\left(
\frac{1+z}{2}\right)^{-1/4}t_2^{1/4}
\end{equation}
\begin{equation}
h \nu_{conf}  = 1 ~ {\rm GeV}~ f_{u,\frac{1}{3}}^2 f_B B_{-5}^3
\left (\frac{E_{54}}{n}\right)^{3/4}\left(\frac{1+z}{2}\right)^{-3/4}t_2^{1/4} ~.
\label{Econf}
\end{equation}
These equations imply that if the observed GeV  photons
were generated by an external shock synchrotron  the upstream
magnetic field must satisfy:
\begin{equation}
 B > 20 ~\mu{\rm  G} \left(\frac{h \nu_{obs}}{10 ~{\rm GeV}}\right)^\frac{1}{3}
 f_{u,\frac{1}{3}}^{-\frac{2}{3}} f_B^{-\frac{1}{3}}\left
(\frac{E_{54}}{n}\right)^{-\frac{1}{4}}\left(\frac{1+z}{2}\right)^{\frac{1}{4}}t_2^{-\frac{1}{12}}
~. \label{Econf}
\end{equation}

Turning now to a wind environment the average Lorentz factor and
blast wave radius that corresponds to an observer time $t$ are
\citep{ChevalierLi00}:
\begin{equation}\label{eqgw}
\Gamma_{wind} = 100 ~ \left(\frac{E_{54}}{A_*}\right)^{1/4}
\left(\frac{1+z}{2}\right)^{1/4} t_2^{-1/4} ~,
 \end{equation}
 and
 \begin{equation}\label{eqrw}
 R_{wind} =
  3.7\times 10^{16}~ {\rm cm} ~ \left(\frac{E_{54}}{A_*}\right)^{1/2}
\left(\frac{1+z}{2}\right)^{-1/2} t_2^{1/2} ~.
  \end{equation}
A substitution of these expression into Eqs. \ref{conf} and
\ref{syn_conf} yields:
 \begin{equation}
  \gamma_{conf-wind}'= 10^8  f_{u,\frac{1}{2}} B_{-5} \left(\frac{E_{54}}{A_*}\right)^{1/2}
\left(\frac{1+z}{2}\right)^{-1/2} t_2^{1/2}
\end{equation}
\begin{equation}
   h \nu_{conf-wind} = 30~ {\rm MeV}~
f_{u,\frac{1}{2}}^2 f_B B_{-5}^3
\left(\frac{E_{54}}{A_*}\right)^\frac{3}{2}
\left(\frac{1+z}{2}\right)^{-\frac{3}{2}} t_2^\frac{1}{2}.
\end{equation}
%
The lower Loretnz factor and the smaller radius (at early time) as
compared to an ISM leads to a more stringent constrain on the
upstream magnetic field of a wind:
\begin{equation}
   B > 70 ~\mu{\rm  G} \left(\frac{h \nu_{obs}}{10 ~{\rm GeV}}\right)^\frac{1}{3}
 f_{u,\frac{1}{2}}^{-\frac{2}{3}} f_B^{-\frac{1}{3}}\left
(\frac{E_{54}}{n}\right)^{-\frac{1}{2}}\left(\frac{1+z}{2}\right)^{\frac{1}{2}}t_2^{-\frac{1}{6}}
~.
\end{equation}

\subsection{Cooling }\label{100MeV}
The conditions found in the previous subsection are sufficient to
produce the highest energy photons. However, the bulk of the energy
typically observed in the LAT comes at lower energies, around 100
MeV. To obtain an efficient,  fast cooling, 100 MeV emission one
needs the cooling frequency, $\nu_c \le  100$ MeV. In both the
upstream and downstream frames the maximal energy, $\nu_{Max}$ is
obtained from the condition $t_{cool} = t_{acc}$, while $\nu_{conf}$
is obtained from the condition $t_{dyn}=t_{acc}$. Since cooling is
more efficient in the downstream, comparing the two conditions in
the downstream frame implies that $\nu_c$ (for which $t_{cool}=
t_{dyn}$) satisfies
\begin{equation}
h \nu_c = h \frac{\nu_{Max}^2}{\nu_{conf,d}}
 \end{equation}
where $\nu_{conf,d} = (4f_d f_B/f_u)^2\nu_{conf}$. This equation of
the cooling frequency is, of course, equivalent to the one obtained
by the traditional derivation \citep[e.g.,][]{SPN98}. Thus, instead
of calculating the exact value of $f_d$ we use the normalization of
$\nu_c$ derived by \cite{GranotSari02} finding:
\begin{equation}
h \nu_c = \left\{\begin{array}{lc}
            60 ~ {\rm GeV}~(f_B B_{-5})^{-3}
\left(\frac{E_{54}}{n}\right)^{-\frac{1}{2}}
\left(\frac{1+z}{2}\right)^{-\frac{1}{2}} t_2^{-\frac{1}{2}}& {\rm ISM} \\
            200  ~{ \rm GeV}~ (f_B B_{-5})^{-3}\left(\frac{E_{54}}{A_*}\right)^{-1}
 t_2^{-1} & {\rm wind}
          \end{array}\right. .
\end{equation}
The wind constrain is again stronger since $\nu_{conf}$ is lower
while $\nu_{Max}$ is independent of the environment. The condition $h
\nu_{c} < 100$MeV, will impose now:
\begin{equation}
f_B B > \left\{\begin{array}{cc}
            85 ~ {\rm \mu G}~
\left(\frac{E_{54}}{n}\right)^{-1/6}
\left(\frac{1+z}{2}\right)^{1/2} t_2^{-1/6}& {\rm ISM} \\
            125  ~{ \rm \mu G}~ \left(\frac{E_{54}}{A_*}\right)^{-1/3}
\left(\frac{1+z}{2}\right)^{2/3} t_2^{-1/3} & {\rm wind}
          \end{array}\right. .
\end{equation}
This is a relatively  high upstream field and it suggests that
either the external density is extremely low or that there is at
least some magnetic field amplification at the shock.

The system has some interesting properties when there is no magnetic
field amplification. In such a case the upstream and downstream
timescales are comparable to within an order of magnitude and each
electron is characterized by only two time scales, its acceleration
time, $t_{acc}$ and its cooling time, $t_{cool}$. It is the interplay
between these two time scales and the dynamical time scale,
$t_{dyn}$, together with the requirement that the electron is bound
to the system, that determine the critical electron Lorentz factors,
$\gamma_{Max}$, $\gamma_{conf}$, and $\gamma_c$. The first satisfies
$t_{acc}(\gamma_{Max})=t_{cool}(\gamma_{Max})$. It is independent of
the system dynamical time and as it turns out $\nu_{Max}$ is also
independent of the strength of the magnetic field (Eq. \ref{Emax}).
Next, equating $t_{acc}=t_{dyn}$ results in the maximal Lorentz
factor an electron can achieve, when cooling and escape are ignored.
In our case, an electron that satisfies $t_{acc}=t_{dyn}$ is also
the maximal energy of an electron that is confined to the system,
thus, $t_{acc}(\gamma_{conf})=t_{dyn}$. Finally, $\gamma_c$ is the
Lorentz factor of the electron that cool over the dynamical time,
i.e., $t_{cool}(\gamma_c)=t_{dyn}$. Therefore there are two
possibilities. If
$t_{acc}(\gamma_{Max})=t_{cool}(\gamma_{Max})<t_{dyn}$ then the
maximal electron's energy is limited by cooling, which takes place
on shorter time than $t_{dyn}$. Thus, less energetic electrons than
$\gamma_{Max}$ can cool over $t_{dyn}$ implying
$\nu_c<\nu_{Max}<\nu_{conf}$. In such case a break in th espectrum
is observed at $\nu_c$ and a cutoff at $\nu_M$, while $\nu_{conf}$
is not observed. On the other hand if
$t_{dyn}<t_{acc}(\gamma_{Max})=t_{cool}(\gamma_{Max})$ then the
maximal electron's energy is limited by the ability to accelerate
and confine the electrons over $t_{dyn}$, where cooling do not play
any role. Thus, $\nu_{conf}<\nu_{Max}<\nu_c$ where $\gamma_{Max}$
and $\gamma_c$ electrons do not exist in the system (cannot be
accelerated). Therefore a cutoff is observed at $\nu_{conf}$ and
$\nu_c$ and $\nu_{Max}$ are not observed. For an ISM there is a
triple coincidence of the three energies for our canonical
parameters when $ B=20 \mu$G so $\nu_{conf}=\nu_{Max}=\nu_c$.

\section{Inverse Compton}

So far we have ignored the possible effects of Inverse Compton (IC)
cooling on the synchrotron emitting electrons. However, the strong
lower energy radiation fields may lead to IC cooling whose effect we
consider now. This radiation field may be the external forward shock
synchrotron field, but it may also be any other source. The IC
cooling may reduce the maximal energy bellow maximal synchrotron
energy (Eq. \ref{Emax}). It leads to an implicit equation for the
new (lower) maximal energy:
\begin{equation}
h \nu_{_{M,IC}} =  \frac{m_e c^2}{\alpha [1+Y(\gamma_{_{M,IC}})]},
\label{EmaxIC} ~,
\end{equation}
where $Y$ is the Compton parameter in the downstream (where the
dominant cooling takes place) and  $\gamma_{_{M,IC}}$ is the Lorentz
factor of the electrons whose synchrotron frequency is
$\nu_{_{M,IC}}$. Clearly, for efficient GeV emission
$Y(\gamma_{_{M,IC}})$ must be smaller than unity and in this case
$\gamma_{_{M,IC}} \approx \gamma_{Max}$ and we recover Eq. \ref{Emax}.
In the following we assume that $Y < 1$ and $\gamma_{_{M,IC}}
\approx \gamma_{Max}$ and consider what are the conditions for this to
hold.

As $\gamma_{Max}$ is extremely high  IC scattering is mostly in the
Klein Nishina (KN) regime and is ineffective. Still, the radiation
field (in the shock frame),  $U'_{rad}$, below the relevant  KN
frequency, $\nu_{_{KN}}(\gamma) $:
\begin{equation}
h \nu_{_{KN}}(\gamma) \equiv \frac{m_e c^2 \Gamma}{\gamma (1+z)}~,
\end{equation}
may be large enough leading to an effective   Compton parameter:
\begin{equation}
Y(\gamma_{Max}) = \frac{U'_{rad}(\nu < \nu_{_{KN}}(\gamma_{Max}))}{U'_B} ~.
\end{equation}
One can estimate $U'_{rad}$ using a given model for the external
shock emission, but given the multi wavelength observations at early
time it is much more useful to constrain it directly from the
observations. This is especially important since, at least in the
earlier part of the afterglow, other radiation fields (e.g. prompt
emission and reverse shock emission) co-exist with the external
shock and this is the best way to incorporate their contributions.
Given an observed spectral flux density $F_\nu$
\begin{equation}
U'_{rad}(\nu < \nu_{_{KN}})  = \max\{ \nu F_\nu(\nu < \nu_{_{KN}})\}
\frac{ d^2 ~t (1+z)}{R^3}   ,
\end{equation}
where  $d$ is the comoving distance. In the following we measure
$F_\nu$ in units of mJy (i.e., $F_\nu = F_{26}
10^{-26}$ergs/cm$^2$Hz). $\max\{ \nu F_\nu(\nu < \nu_{_{KN}})\}$ is
the maximal value of $\nu F_\nu$ at frequencies below $\nu_{_{KN}}$.
Below we assume that this maximum is at $\nu_{_{KN}}$, an assumption
that can be tested, and corrected, given the observations.

Using Eqs. \ref{eqgs}, \ref{eqrs}, \ref{eqgw} and \ref{eqrw}  we
find that $\nu_{_{KN}}(\gamma_{Max})$ is typically in the IR when there
is no strong field amplification:
\begin{equation}
\nu_{_{KN}}(\gamma_{Max}) = \left\{\begin{array}{cc} 0.05~ {\rm
eV}\sqrt{{f_B B_{-5}}} \left(\frac{E_{54}}{n}\right)^{\frac{3}{16}}
\left(\frac{1+z}{2}\right)^{\frac{-7}{16}} t_2^{-\frac{9}{16}} &{\rm ISM,} \\
 0.02~ {\rm eV}  \sqrt{{f_B B_{-5}}} \left(\frac{E_{54}}{A_*}\right)^{\frac{3}{8}}
\left(\frac{1+z}{2}\right)^{-\frac{5}{8}} t_2^{-\frac{3}{8}}& {\rm
wind.}
          \end{array}\right. .
\end{equation}
The relevant Y parameter assuming $\nu F_\nu(\leq \nu_{_{KN}})$
peaks at $\nu_{_{KN}}$   is:
\begin{equation}
Y(\gamma_{Max})=\left\{\begin{array}{cc}
 0.2~  {F_{26}} {d_{28}}^{2} (f_B B_{-5})^{-\frac{3}{2}}\left(\frac{E_{54}}{n}\right)^{-\frac{13}{16}}
\left(\frac{1+z}{2}\right)^{\frac{9}{16}} t_2^{\frac{7}{16}}
 & {\rm ISM,} \\
30 (f_B B_{-5})^{-\frac{3}{2}}
\left(\frac{E_{54}}{A_*}\right)^{-\frac{13}{8}}
\left(\frac{1+z}{2}\right)^{\frac{11}{8}} t_2^{-\frac{3}{8}}
   & {\rm  wind.}
\end{array}\right.
\end{equation}
Additionally, in order to produce efficiently the emission observed
by LAT at $>100$ MeV,  the corresponding electrons cannot be cooled
effectively via IC, and  the Compton parameter should  be small. The
relevant KN frequency is:
\begin{equation}
\begin{array}{ll}
  \nu_{_{KN}}(100{\rm ~MeV}) =&  \\
  \left\{\begin{array}{ll}
 0.5  ~{\rm eV} ~ (f_B B_{-5}) ^\frac{1}{2} \left(\frac{E_{54}} {n}\right)^\frac{1}{4}
 \left( \frac{1+z}{2} \right)^{-\frac{3}{4}} {t_2}^{-\frac{3}{4}} & {\rm ISM,} \\
 0.2 ~{\rm eV} ~(f_B B_{-5} )^\frac{1}{2} \left ( \frac {E_{54}} { A_*}\right)^\frac{1}{2}
 \left(\frac{1+z}{2}\right)^{-1}  t_2^{-\frac{1}{2}}
& {\rm  wind.}
\end{array}\right.&
\end{array}
\end{equation}
and assuming $\nu F_\nu(\leq \nu_{_{KN}})$ peaks at $\nu_{_{KN}}$
the Compton $Y$  is:
\begin{equation}
\begin{array}{ll}
Y(100 {\rm ~MeV})=\\
 \left\{\begin{array}{cc}
 1.3  {F_{26}} {d_{28}}^{2} (f_B B_{-5})^{-\frac{3}{2}} \left(\frac{E_{54}}{n} \right)^{-\frac{3}{4}}   \left(
 \frac{1+z}{2} \right)^\frac{1}{4} t_2^\frac{1}{4}  & {\rm ISM,}  \\
180 ~ {F_{26}} {d_{28}}^{2} (f_B B_{-5})^{-\frac{3}{2}} \left(\frac{ E_{54}}{ A_*}\right)^{-\frac{3}{2}}
  \frac{1+z}{2} t_2^{-\frac{1}{2}}
 & {\rm  wind.}
\end{array}\right.
\end{array}
\end{equation}
Therefore, if there is no strong field amplification a modest
IR-Optical flux of a few mJy for an ISM and 10 $\mu$Jy for a wind is
enough to suppress the GeV flux. Therefore a clear test of the
external shock synchrotron GeV emission, without field amplification
is that we should not observe a strong optical flux at the time when
we observe GeV emission. A simultaneous detection of a strong
optical emission and a GeV emission will rule out this model.

Although we do not consider here a complete model for the external
shock synchrotron emission it is important to note that when such
model is constructed KN effects should be carefully considered. The
reason is that if the magnetic field is not amplified close to
equipartition level then the radiation energy density is much larger
than the magnetic field energy density in the downstream frame. IC
by electrons emitting the MeV-GeV emission will certainly be in the
KN regime, but over a large range of the parameter phase the Y
parameter of these fast cooling electrons (which now depends on the
electron energy) will be larger than 1. In which case the
synchrotron spectrum will be altered significantly \citep{Nakar09}.
An extreme example to a KN effect may be apparent if there is no
field amplification at all. In such case $\nu_{_{KN}}$ of the fast
cooling electrons is below the the typical synchrotron frequency
$\nu_{Max}$.If their Compton parameter is smaller than unity and the
synchrotron emission is not strongly affected. But, once $Y$ becomes
larger than $1$ for fast cooling electrons, the back-reaction of the
IC emission on the electrons distribution results in the cooling
frequency "jumping" on a short time scale by orders of magnitude,
significantly revising the whole synchrotron spectrum
\citep{Nakar09}.

\section{Conclusions}

Using the confinement and cooling conditions we have obtained limits
on the values of the magnetic fields needed in the downstream and
upstream regions in order to produce the observed GRB GeV emission
via an external shock synchrotron.  These constrains are based on
minimal assumptions of synchrotron cooling and blast wave
hydrodynamics. Both are essential ingredients of the external shock
synchrotron model.  The arguments we present allow us to explore
directly the magnetic fields in both upstream and downstream
regions, which are among the least constrained physical
parameters of the model.

We find that with no amplification the minimal fields required are
on the high end  ($\tilde100 \mu$ G), unless the external density is
very low. The limits are even higher for a radiative solution. It
is, of course, possible that this is a condition for GeV emission.
However, the detection of GeV emission from all MeV bright GRBs that
are within the LAT viewing angle suggests that the emission is
generic. In this case at least a modest amplification is
probably needed.

Finally, we point out two critical predictions of the external shock
synchrotron model: (i)  No detection of late very energetic ($>10$
GeV) photons and (ii) No simultaneous detection of a  bright
($>$mJy) IR-optical (depending on the specific case) signal with the
GeV photons unless the upstream magnetic field is strongly
amplified in the shock. Continued observations should be compared
with these predictions and can provide future tests of this model.

This research was supported by an ERC advanced research grant, by
the ISF center for excellence for high energy astrophysics, ISF
grant No. 174/08,
and  by an IRG Marie-Curie
Grant. We thank Pawan Kumar and Rudolfo Barniol Duran and Boaz Katz
for helpful discussions.


\end{document}